\documentclass[aps,prl,twocolumn,superscriptaddress]{revtex4-1}

\usepackage[utf8]{inputenc}
\usepackage[T1]{fontenc}
\usepackage{amsmath}
\usepackage{amssymb}
\usepackage{amsfonts}
\usepackage{bm}
\usepackage{color}
\usepackage{datetime}
\usepackage{graphicx}
\usepackage[colorlinks,citecolor=blue,urlcolor=blue,linkcolor=blue]{hyperref}
\usepackage{physics}
\usepackage{mathtools}
\usepackage{enumerate}

\DeclareMathSymbol{*}{\mathbin}{symbols}{"01}
\newcommand{\inext}{{}_\mathrm{ext}}
\newcommand{\inint}{{}_\mathrm{int}}
\newcommand{\expv}[1]{\expval*{#1}}
\newcommand{\SN}[1][N]{\text{S}_{#1}}
\DeclareMathOperator{\Var}{Var}

\begin{document}

\title{Many-body interference at the onset of chaos}

\author{Eric Brunner}
\email{ecj.brunner@gmail.com}
\affiliation{\freiburg}
\affiliation{\eucor}
\author{Lukas Pausch}
\altaffiliation[Current address: ]{CESAM Research Unit, University of Liège, 4000 Liège, Belgium}
\affiliation{\freiburg}
\affiliation{\eucor}
\author{Edoardo G. Carnio}
\affiliation{\freiburg}
\affiliation{\eucor}
\author{Gabriel Dufour}
\affiliation{\freiburg}
\affiliation{\eucor}
\author{Alberto Rodr\'{\i}guez}
\affiliation{\salamanca}
\affiliation{\salamancaii}
\author{Andreas Buchleitner}
\affiliation{\freiburg}
\affiliation{\eucor}

\newcommand{\freiburg}{Physikalisches Institut, Albert-Ludwigs-Universit\"{a}t Freiburg, Hermann-Herder-Stra{\ss}e 3, 79104, Freiburg, Germany}
\newcommand{\eucor}{EUCOR Centre for Quantum Science and Quantum Computing, Albert-Ludwigs-Universit\"{a}t Freiburg, Hermann-Herder-Stra{\ss}e 3, 79104, Freiburg, Germany}
\newcommand{\salamanca}{Departamento de F\'isica Fundamental, Universidad de Salamanca, E-37008 Salamanca, Spain}
\newcommand{\salamancaii}{Instituto Universitario de Física Fundamental y Matemáticas (IUFFyM), Universidad de Salamanca, E-37008 Salamanca, Spain}


\begin{abstract}

We unveil the signature of many-body interference across dynamical regimes of the Bose-Hubbard model. Increasing the particles' indistinguishability enhances the temporal fluctuations of few-body observables, with a dramatic amplification at the onset of quantum chaos. By resolving the exchange symmetries of partially distinguishable particles, we explain this amplification as the fingerprint of the initial state's coherences in the eigenbasis.
\end{abstract}

\maketitle

Interacting many-particle dynamics may be considered {\em the} most 
plausible origin of instabilities, chaos and complexity,
from astronomical \cite{chirikov_1989,laskar_2009} to microscopic \cite{Bohr1936} scales. Due to the rapid growth of phase space with the particle number, together 
with its progressively 
more intricate topology, deterministic descriptions quickly hit the ceiling, enforcing statistical
descriptions. 
Some type of coarse graining, implicit to such approaches, allows classifications of dynamical behavior---e.g.,
as scale-invariant, chaotic or Markovian---associated with {\em universal} characteristics which are formalized, e.g., in the theories of phase transitions \cite{landau1984,sachdev2011}, random matrices (RMT) \cite{Mehta1991}, or open quantum systems \cite{davies1976,alicki1987,breuer2002}. It is the 
universal character of these features which 
allows robust predictions, since full resolution of complex dynamics is prohibitive, by their very nature. 

On the quantum level, robust 
features are in such scenarios essentially controlled by spectral densities and statistics, the localization properties of eigen- and initial states, the phase-space dimension, and the time scales over which to make predictions. This is the unifying view of quantum chaos \cite{LesHouches}, which has proven enormously versatile an approach to analyze complex quantum systems---including paradigmatic many-particle 
scenarios in nuclear \cite{guhr1989,rotter_1991} and atomic physics \cite{Holle1988,iu1991,krug_2001}, as 
well as in cold matter \cite{moore_1995,wimberger2004,Meinert2014,Kaufman2016} and 
black hole \cite{liu_2020} contexts.
On this level of description, the specific many-particle nature of the underlying Hamiltonian does not appear as an essential ingredient anymore, since all the features of complex dynamics can also be observed on the level of single-particle dynamics \cite{LesHouches,garbaczewski_2002} (provided the phase space dimension is large enough---such that tori are not isolating anymore \cite{uzer_1996}).  

Yet, quantum systems composed of identical particles undeniably exhibit properties that fundamentally distinguish them from classical many- and 
single-particle systems, 
hardwired in exchange symmetries \cite{landau1985,Tichy2017}, and generating many-body interference (MBI) phenomena
\cite{HOM1987,Tichy2010,Dufour2017,shchesnovich_collective_2018,Giordani2018,Dittel2018,Dittel2018b,brunner_signatures_2018,jones_multiparticle_2020,dufour_many-body_2020,dittel_wave-particle_2021,brunner_many-body_2022,Tichy2012,stanisic_discriminating_2018,dufour_many-body_2020,dittel_wave-particle_2021}, thus with potentially dramatic dynamical relevance. 
In fact, modern experiments \cite{Greiner2002,Bloch2008,Gericke2008,Gemelke2009,Karski2009,Bakr2010,Sherson2010,Kaufman2016,Lukin2019,Rispoli2019} already
allow to control external and internal degrees of freedom (dof) of many-particle quantum systems, such that physically identical 
particles may be equipped with a continuously tunable degree of partial distinguishability (PD), and, by this, to ultimately control
the impact of MBI on the dynamics \cite{brunner_signatures_2018,dufour_many-body_2020,dittel_wave-particle_2021,brunner_many-body_2022}.
While, 
traditionally, the RMT approach 
deliberately divides out 
any symmetry-induced properties \cite{LesHouches,Mehta1991} (see, however, \cite{giraud2022}), 
it is clear that MBI, as a manifestation of the specific system's particle exchange symmetry, is one of those robust features which need to be accounted for in any theory of complex quantum systems. 
This raises the question:
{\em Where in a many-body quantum system's spectral and eigenstate structure is MBI encoded, and how can we distill its impact 
on observable dynamical properties?}

\begin{figure}
\centering
\includegraphics[width=0.48\textwidth]{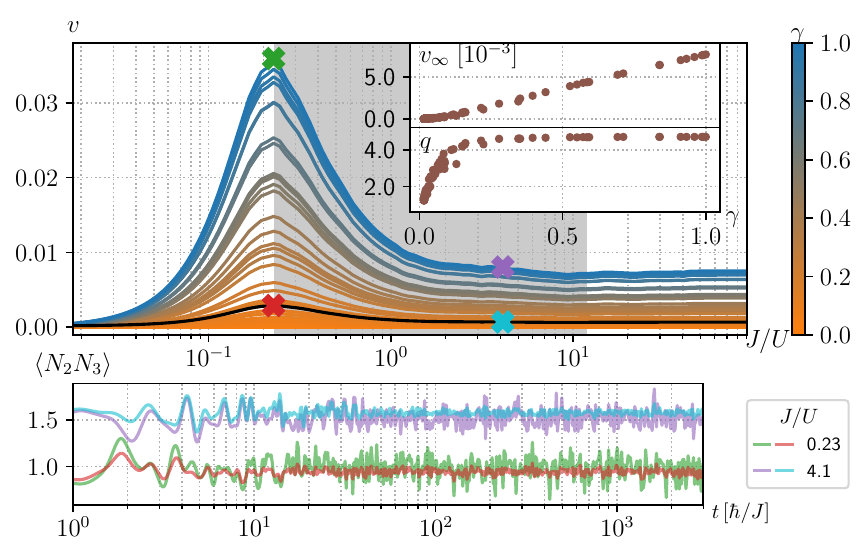}
\caption{Fluctuations of two-particle observables in the Bose-Hubbard model \eqref{eq:BHM} for $N=L=6$. 
Top: Average temporal variance $v$ [Eq.~\eqref{eq:time_var}] vs.\ $J/U$, for 100 initial states with one particle per site and
variable particle indistinguishability $\gamma$ (color bar). The black curve highlights the case analyzed in Fig.~\ref{fig:vard1_vs_sectors}(b).
The grey-shaded area indicates the quantum-chaotic region of \eqref{eq:BHM} (also see Figs.~\ref{fig:vard1_vs_sectors}(b,c)).
Insets show the plateau value $v_\infty$ (large $J/U$) and the enhancement $q = v_\text{max}/v_{\infty}$ of the maximal fluctuation $v_\text{max}$, as functions of $\gamma$.
Bottom: Time series of $\langle N_2(t)N_3(t)\rangle$ for four combinations of $J/U$ and $\gamma$ (identified by correspondingly colored crosses in the top panel). 
For visibility, the curves for $J/U=4.1$ are shifted upwards by 0.8.}
\label{fig:vk_vs_nu_pd}
\end{figure}

In this contribution, we identify a signature of bosonic MBI in the asymptotic temporal fluctuations $v$ of expectation values around their average.
We show that $v$ is controlled by the coherences of the many-particle initial state in the eigenbasis, multiplied by the corresponding off-diagonal elements of the observable, and is therefore strongly enhanced by particle indistinguishability. 
We extract $v$ from the quench dynamics of a Mott state in the Bose-Hubbard model for increasing values of tunnelling strength $J$.
As shown in Fig.~\ref{fig:vk_vs_nu_pd}, $v$ is sharply peaked around the value of $J$ where the dynamics becomes chaotic. 
There, the initial state is sufficiently delocalized in the eigenbasis for coherences to build up, but not so much that they are cut off by the finite energy bandwidth of the observable.
By taking into account the eigenstates' structure as constrained by their symmetry under particle exchange,
we find the strongest dependence of $v$ on the particles' mutual (in)distinguishability precisely at that point.
Given the very general ingredients of our theoretical analysis, we conclude that many-body coherence effects are most intense at the onset of quantum chaos.

We consider the one-dimensional  Bose-Hubbard model \cite{Fisher1989,Lewenstein2007,Bloch2008,Cazalilla2011,Krutitsky2016} of PD particles with hard-wall boundary conditions, 
\begin{equation}\label{eq:BHM}
	H= -J \sum_{\langle i, j\rangle} \sum_{\sigma=1}^s  a^\dagger_{i\sigma} a_{j \sigma} + \frac{U}{2} \sum_{i=1}^L N_i(N_i -1) \,,
\end{equation}
which is experimentally realizable with ultracold atoms in optical lattices \cite{Greiner2002,Bloch2008,Gericke2008,Gemelke2009,Karski2009,Bakr2010,Sherson2010,Kaufman2016,Lukin2019,Rispoli2019}.
The first index of the creation and annihilation operators $a^\dagger_{i\sigma}, a_{j\sigma}$ refers to the $L$ 
Wannier orbitals of the 
lattice, which span the \emph{external} single-particle Hilbert space $\mathcal{H}\inext$. 
The second index $\sigma$ refers to a basis of the $s$-dimensional \emph{internal} single-particle Hilbert space, describing, e.g., the electronic state of an atom 
loaded into an optical lattice.
The operator $N_i = \sum_{\sigma=1}^s a^\dagger_{i\sigma}a_{i\sigma}$ counts the number of particles on lattice site $i$, irrespective of their internal state. We keep the total 
particle number $N=\sum_{i=1}^L N_i$ fixed.
The two terms in $H$ describe nearest-neighbor 
tunneling and on-site interaction of the particles, both of which act exclusively on the external dof, while the
internal dof remain static.
For indistinguishable bosons, depending on the relative contribution of both terms in \eqref{eq:BHM}, a quantum-chaotic region has been identified both from spectral statistics and eigenstate delocalization \cite{Buchleitner2003,Kolovsky2004,Kollath2010,Beugeling2015c,Dubertrand2016,Kaufman2016,Beugeling2018,Lukin2019,Rispoli2019,pausch_chaos_2021,pausch_chaos_2021_2,pausch_chaos_2021_3}. We here establish its existence also for PD particles, as an important corollary of our subsequent analysis.

Of experimental interest are few-particle observables, e.g., low-order density correlations $O = N_i N_j$.
Formally, these are given by 
products of $k$ creation and $k$ annihilation operators \cite{brunner_signatures_2018,dufour_many-body_2020}, $k\ll N$, such that they only access the marginal information inscribed in the 
$k$-particle ($k$P) reduced state \cite{brunner_many-body_2019,brunner_many-body_2022}.
Moreover, like the Hamiltonian, these observables are assumed to exclusively act on external dof, such that we can consider their restriction to $\mathcal{H}\inext^{\otimes N}$ and trace out the internal dof from the full system state $\varrho$ to obtain $\rho = \text{tr}\inint \varrho$ \cite{brunner_many-body_2019,dittel_wave-particle_2021,brunner_many-body_2022}.
Partial distinguishability of the particles results in entanglement between their external and internal dof \cite{brunner_many-body_2022,dittel_wave-particle_2021}, 
and we use as a measure of indistinguishability the purity $\gamma = \text{tr} \rho^2$ of the external state, which is maximal ($\gamma=1$) for indistinguishable particles, and  minimal for perfectly distinguishable ones \cite{brunner_many-body_2019,dittel_wave-particle_2021}.

The system's dynamical equilibration, on asymptotic time scales, is captured by the temporal variance of expectation values $\expv{O(t)}$:  
\begin{equation}\label{eq:temp-var}
\text{Var}_t[O] = \overline{\expv{O(t)}^2} - \overline{\expv{O(t)}}^2 \,,
\end{equation}
where $\overline{\phantom{i}\ldots \phantom{i}}$ indicates the average over the positive time axis \cite{SM}.
To formulate general statements, independent of 
the specific choice of observable $O$, we consider an unbiased average (indicated by $\widehat{\phantom{i}\dots \phantom{i}}$) over an orthonormal basis $\mathcal{B}$ of the (finite-dimensional) Hilbert space of external $k$P 
observables \cite{N1},
\begin{equation}\label{eq:time_var}
	v  \coloneqq  \widehat{\text{Var}_t[O] } = \sum_{o\in\mathcal{B}} \text{Var}_t[o] \,,
\end{equation} 
This quantity is shown, for $k=2$, in Fig.~\ref{fig:vk_vs_nu_pd} (top panel), for the dynamics generated by \eqref{eq:BHM}, 
with $N=L=6$, initially one particle per external mode, versus the control parameter $J/U$.
A variable level of PD
is obtained by random generation \cite{SM} of the particles' internal states $\ket{\phi_i} = \sum_{\sigma}\phi_{i\sigma} \ket{\sigma}$, $i = 1,\dots, L$, of the initial Mott state, such as to smoothly cover the entire range $\gamma\in [1/N!;1]$.

\begin{figure*}
\centering
\includegraphics[width=\textwidth]{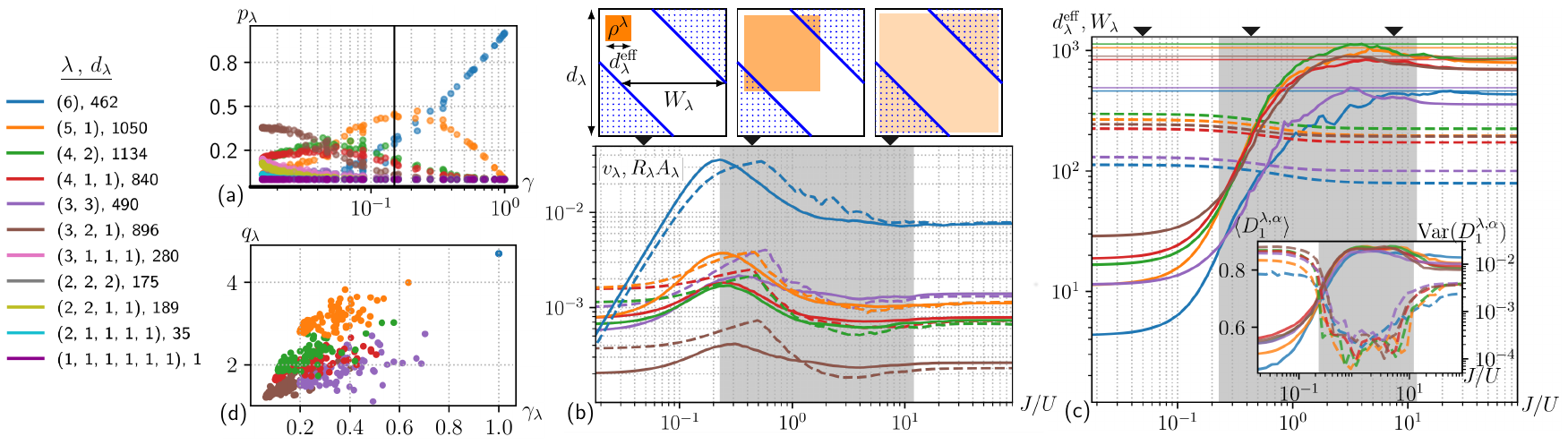}
\caption{
Ingredients determining the averaged temporal variance $v$ shown in Fig.~\ref{fig:vk_vs_nu_pd}, as functions of $J/U$ and of the particles' indistinguishability $\gamma$. 
The legend color-codes the symmetry sectors $\lambda$
and indicates
their corresponding block dimensions $d_\lambda$. 
(a) Weights $p_\lambda$ [see Eq.~\eqref{eq:vk}] of the initial states for varying $\gamma$ (vertical black line marks $\gamma = 0.15$). 
(b) Bottom panel: Comparison of $v_\lambda$ (solid), Eq.~\eqref{eq:vk}, to the factorized approximation $R_\lambda A_\lambda$ (dashed), Eq.~\eqref{eq:fact}, versus $J/U$, for $\gamma\simeq 0.15$.
Top panel:
Qualitative illustration of the competition between the delocalization $d_\lambda^\mathrm{eff}$ of the initial state's component $\rho^\lambda$ and the bandwidth $W_\lambda$ of the averaged observable $\widehat{| O^\lambda_{\alpha\beta} |^2}$, in the eigenbasis of the symmetry sector $\lambda$, for three values of $J/U$.
(c) Evolution of $W_\lambda$ (dashed) and $d_\lambda^\mathrm{eff}$ (solid) with $J/U$ [horizontal lines highlight block dimensions $d_\lambda$]. 
Inset: Delocalization of the $60$ eigenstates closest in energy to the initial state, quantified by the mean value (solid) and variance (dashed) of their fractal dimensions with respect to the Young basis, identifying the chaotic phase (shaded regions in the inset and in the main panels (b,c)). 
(d) Distribution of the sector-specific enhancement $q_\lambda$ for initial states with variable $\gamma_\lambda$. For clarity, panels (b) to (d) show data for the largest six sectors only.
}
\label{fig:vard1_vs_sectors}
\end{figure*}

We observe that, for all $J/U$, $v$ monotonically grows with $\gamma$, i.e., as MBI contributions are enhanced.
Moreover, $v$ exhibits a maximum $v_\text{max}$ at $J/U \simeq 0.23 $, and then decreases to a plateau value $v_{\infty}$ with increasing $J/U$.
The peak is located at the transition to the (grey shaded) parameter range where \eqref{eq:BHM} exhibits fully developed quantum chaos, as identified by the ergodicity properties of its eigenstates (see discussion of Fig.~\ref{fig:vard1_vs_sectors}(c) below).
Both $v_{\infty}$ and the enhancement $q = v_\text{max}/v_{\infty}$ of the fluctuations at the peak increase monotonically with $\gamma$, as shown in the inset. In particular, $q$ steeply increases at small $\gamma$, when MBI starts to kick in, which signals a particularly strong sensitivity to MBI at the transition to quantum chaos.
We observe the same qualitative behavior (see \cite{SM}) for the experimentally more accessible average $\sum_{i \neq j} \text{Var}_t[N_i N_j]$ 
over all two-point density correlations \cite{Giordani2018,walschaers2016}. In the bottom panel of Fig.~\ref{fig:vk_vs_nu_pd}, we also give the long-time series of $\expv{N_2(t)N_3(t)}$, for four values of $J/U$ and $\gamma$, with strongest 
fluctuations for intermediate $J/U \simeq 0.23$ and $\gamma = 1$, in agreement with the above. 
The peak in the fluctuations at the onset of chaos can be qualitatively explained by the competition between the initial state's delocalization and the observable's bandwidth in the eigenbasis, as sketched in the top panels of Fig.~\ref{fig:vard1_vs_sectors}(b).
However, a precise discussion requires to first consider the particle-exchange symmetry of PD bosons, which is at the origin of the overall increase of $v$ with indistinguishability $\gamma$.

States of PD bosons are characterized by the coexistence of several types of mixed
particle-exchange symmetries, alongside the fully symmetric, bosonic symmetry
\cite{Tichy2017,dufour_many-body_2020}.	
The suppression of $v$ as we make particles more distinguishable (i.e., for decreasing $\gamma$) can be understood by the emergence of such	non-bosonic contributions to the dynamics. 
Indeed, group representation theory, and specifically the Schur-Weyl duality \cite{Fulton2004,rowe_dual_2012}, tell us that
$H$, $O$ and $\rho$ (as operators on $\mathcal{H}\inext^{\otimes N}$)
decompose into symmetry sectors labelled by the integer partitions of $N$ (or Young diagrams): $\lambda=(N), (N-1,1), (N-2,2),  \dots, (1,1,\dots)$.
While states $\rho$ of perfectly indistinguishable bosons are entirely supported on the \emph{bosonic} sector, $\lambda = (N)$, states of PD particles also have finite weights $p_\lambda$ on the other sectors ($\sum_\lambda p_\lambda = 1$), as shown in Fig.~\ref{fig:vard1_vs_sectors}(a) for states with variable levels of indistinguishability $\gamma$ (as used as initial states in Fig.~\ref{fig:vk_vs_nu_pd}).
Every sector further decomposes  \cite{SM} into $\nu_\lambda$ identical blocks, each of dimension $d_\lambda$, and we denote by $\{\ket{\lambda, m}\}_{m=1,\ldots,d_\lambda}$ the Young basis \cite{Fulton2004,dufour_many-body_2020}, built upon the Wannier basis, of one such block.
Diagonalizing $H$ in this very block, we find the eigenstates 
$\ket{E^\lambda_\alpha} = \sum_{m=1}^{d_\lambda} c^\lambda_{\alpha m} \ket{\lambda, m }$, with respect to which we represent the observable and the initial state, $O^\lambda_{\alpha\beta} = \expv{E^\lambda_\alpha | O | E^\lambda_\beta}$ and $\rho^\lambda_{\alpha\beta} = \nu_\lambda \expv{E^\lambda_\alpha | \rho | E^\lambda_\beta}/p_\lambda$.
With this definition, the matrix $\rho^\lambda$ has unit trace. If we denote its purity by $\gamma_\lambda = \text{tr}\, [\big(\rho^\lambda \big)^2 ]$, the purity of $\rho$ reads $\gamma = \sum_\lambda p_\lambda^2 \gamma_\lambda/\nu_\lambda$.

The above structure allows to decompose $v$, as given by Eq.~\eqref{eq:time_var}, into contributions from individual symmetry sectors. In the absence of degeneracies between energy levels and between energy gaps, 
within each block and between $\lambda$-sectors, we obtain \cite{SM}
\begin{equation}\label{eq:vk}
v  = \sum_\lambda p_\lambda^2 \, v_\lambda \,, \quad v_\lambda = \sum_{\alpha \neq \beta} |\rho^\lambda_{\alpha\beta}|^2 \widehat{|O^\lambda_{\alpha\beta} |^2} \,.
\end{equation}
The individual contributions are determined by the \emph{squared} weights $p_\lambda^2$, and by the \emph{off-diagonal} elements $|\rho^\lambda_{\alpha\beta}|^2, \widehat{|O^\lambda_{\alpha\beta}|^2}$ of initial state and observable in the eigenbasis.
For indistinguishable particles, $\gamma=1$ and $p_\lambda=\delta_{\lambda,(N)}$, the fluctuations $v=v_{(N)}$ are thus governed by purely bosonic MBI.
As $\gamma$ decreases, the state starts to distribute over other sectors, as shown in Fig.~\ref{fig:vard1_vs_sectors}(a). The fluctuations $v$ are then doubly suppressed: through the squared weights $p_\lambda^2$ in Eq.~\eqref{eq:vk}, and because of $v_\lambda\leq v_{(N)}$ for all $\lambda$, as we will show in Fig.~\ref{fig:vard1_vs_sectors}(b).
In the limit of distinguishable particles (smallest $\gamma =1/N!$), the state is distributed over all sectors and $v$ is minimal.

To elucidate the origin of the maximum of $v$ at the transition to quantum chaos, we develop a simple statistical model for the off-diagonal elements of state and observable appearing in Eq.~\eqref{eq:vk}.
We assume that the averaged matrix elements $\widehat{| O^\lambda_{\alpha\beta} |^2}$ vanish outside a band of width $W_\lambda$ \cite{N2,SM},  as suggested by the eigenstate thermalization hypothesis \cite{feingold_distribution_1986,Deutsch1991,Srednicki1994,Srednicki1999,hiller_complexity_2006,deutsch_eigenstate_2018}.
As for the state $\rho^\lambda$, we suppose that it only populates $d_\lambda^\mathrm{eff}$ consecutive (in energy) eigenstates, as sketched in the top panels of Fig.~\ref{fig:vard1_vs_sectors}(b).
Otherwise,  $| \rho^\lambda_{\alpha\beta}|^2$ and $\widehat{| O^\lambda_{\alpha\beta}|^2}$ are assumed to be statistically independent, such that we can factorize (see \cite{SM})
\begin{equation}\label{eq:fact}
	v_\lambda \approx R_\lambda A_\lambda\, , R_\lambda  = \frac{\sum_{\alpha \neq\beta} | \rho^\lambda_{\alpha\beta} |^2}{\max(d_\lambda^\mathrm{eff}, W_\lambda)} \, , A_\lambda =  \frac{\sum_{\alpha \neq\beta} \widehat{| O^\lambda_{\alpha\beta} |^2} }{d_\lambda} \,.
\end{equation}
This is qualitatively underpinned by Fig.~\ref{fig:vard1_vs_sectors}(b), for a state with $\gamma = 0.15$, for the largest six symmetry sectors (which carry $99.2\%$ of $\rho$).
Note that weighting those $v_\lambda$ by the squares of the associated $p_\lambda$ [black vertical line in Fig.~\ref{fig:vard1_vs_sectors}(a)], according to Eq. \eqref{eq:vk}, results in the black curve highlighted in the upper panel of 
Fig.~\ref{fig:vk_vs_nu_pd}.

We find that $A_\lambda$ is, to a good approximation, independent of $J/U$, and of order one in all contributing sectors \cite{N3,SM}.
Consequently, the dependence of $v_\lambda$ on $J/U$ is predominantly controlled by $R_\lambda$.
From $R_\lambda$ [cf. Eq.~\eqref{eq:fact}] we rewrite the sum over coherences  $\sum_{\alpha \neq\beta} | \rho^\lambda_{\alpha\beta} |^2=\gamma_\lambda-I_\lambda$, where the inverse participation ratio $I_\lambda = \sum_\alpha | \rho^\lambda_{\alpha\alpha} |^2$ is a measure of the initial state's localization in the eigenbasis, which we now turn to.

Since $I_\lambda^{-1}$ provides an estimate of the number of eigenstates occupied by the initial state, we use it to define the state's effective dimension $d_\lambda^\mathrm{eff} = C_\lambda/I_\lambda$.
The multiplicative factor $C_\lambda$ enforces $d_\lambda^\mathrm{eff} \rightarrow d_\lambda$ in the regime of strongest delocalization, since, due to residual fluctuations of $\rho_{\alpha\alpha}$, $I_\lambda^{-1}$ generically underestimates the actual number of populated eigenstates. 
Figure~\ref{fig:vard1_vs_sectors}(c) illustrates the delocalization of the initial state seeding the fluctuations displayed in Fig.~\ref{fig:vk_vs_nu_pd}, in the energy eigenbasis of the largest six sectors [carrying between $68\%$ (distinguishable particles) and $100\%$ (indistinguishable particles) of the initial state, cf. Fig.~\ref{fig:vard1_vs_sectors}(a)].
From the strongly interacting limit $J/U \rightarrow 0$, $d_\lambda^\mathrm{eff}$ grows with increasing 
tunneling strength, reaching a maximum in most sectors in the range $J/U \in [3;4]$, before stabilizing at a value of order $d_\lambda$ for $J/U\to\infty$.

The delocalization of the initial state in the eigenbasis mirrors the delocalization of the eigenstates in the Young basis $\{\ket{\lambda, m}\}_{m=1,\ldots,d_\lambda}$, which signals
the emergence of quantum chaos.
As demonstrated in Refs.~\cite{pausch_chaos_2021,pausch_chaos_2021_2,pausch_chaos_2021_3} for indistinguishable bosons, 
the chaotic region can be identified by the ergodicity of eigenstates in the individual $\lambda$-sectors, as measured by their
fractal dimension ${D}^{\lambda,\alpha}_1 = - \sum_{m=1}^{d_\lambda} |c^\lambda_{\alpha m}|^2 \log_{d_\lambda} |c^\lambda_{ \alpha m} |^2$.
The inset of Fig.~\ref{fig:vard1_vs_sectors}(c) shows the mean value $\expv{{D}^{\lambda,\alpha}_1}$ and the variance $\Var({D}^{\lambda,\alpha}_1)$, taken over the 60 eigenstates closest in energy to the initial state, for each $\lambda$-sector.
A substantial delocalization occurs for $ 0.23 \lesssim J/U \lesssim 11 $
(shaded area), where the mean values reach their maxima, accompanied by a drop of the variances by at least one order of magnitude, attesting a strongly uniform eigenstate structure in all shown sectors.
Consequently, the chaotic domain identified for indistinguishable bosons \cite{pausch_chaos_2021} persists for mixed particle-exchange symmetry. 

In contrast to the sharp growth of $d_\lambda^\mathrm{eff}$, the bandwidth  $W_\lambda$ of the observable only decreases slightly with $J/U$. We estimate it by taking the standard deviation of the (normalized) distribution $f_\alpha(\beta) \propto  \widehat{ | O^\lambda_{\alpha\beta} |^2}$ for each $\alpha$, and  averaging over $\alpha$.
Figure~\ref{fig:vard1_vs_sectors}(c) shows the resulting $W_\lambda$ (dashed lines) versus $J/U$, in the largest six sectors.

The behavior of $v_\lambda$ can then be qualitatively understood in terms of the three regimes sketched at the top of Fig.~\ref{fig:vard1_vs_sectors}(b).
In the limit $J/U \rightarrow 0$ (leftmost sketch), the initial Mott state is itself an eigenstate and decomposes on only a few (degenerate) eigenstates $\ket{E_\alpha^\lambda}$. Accordingly, the inverse participation ratio is maximal, yielding a minimal value of the sum over coherences $\gamma_\lambda-I_\lambda$, as captured by the decreasing left tails of the $v_\lambda$ in Fig.~\ref{fig:vard1_vs_sectors}(b). 
Instead, for $J/U$ within and beyond the range of fully developed quantum chaos (rightmost sketch), the initial state is strongly delocalized in the eigenbasis ($d_\lambda^\mathrm{eff}\approx d_\lambda$),
such that many non-zero coherences $|\rho^\lambda_{\alpha\beta}|^2$ with $|\alpha-\beta|>W^\lambda$ are suppressed by multiplication with a vanishing $\widehat{|O^\lambda_{\alpha\beta}|^2}$ in Eq.~\eqref{eq:vk}. In the factorized form Eq.~\eqref{eq:fact}, this effect gives rise to the denominator $\max(d_\lambda^\mathrm{eff}, W_\lambda)$ of $R_\lambda$,
which results
(with $I_\lambda \ll \gamma_\lambda$, $d_\lambda^\mathrm{eff} \approx d_\lambda > W_\lambda$)
in a small asymptotic value $R_\lambda^\infty \approx \gamma_\lambda/d_\lambda \ll 1$ for large $J/U$.
At the transition between the two parameter ranges (central sketch),
the onset of quantum chaos, 
where the eigenstates undergo a metamorphosis
from localized to ergodic,
triggers the initial state's delocalization in the eigenbasis.
There, $R_\lambda$ exhibits a maximum, since $\rho^\lambda$ already populates a substantial energy window, resulting in an enhanced contribution by coherences, which are, however, not yet suppressed by the observable's  bandwidth.

To explain why the effect of PD on $v$ is comparatively strongest at this maximum, we turn to the dependence of $v_\lambda$ on the purity $\gamma_\lambda$ of the state $\rho^\lambda$ associated with a given symmetry sector: We have seen that the plateau value $v_\lambda^\infty\sim R_\lambda^\infty$ scales linearly with $\gamma_\lambda$. 
In Fig.~\ref{fig:vard1_vs_sectors}(d), we observe a correlation of the sector-specific enhancement $q_\lambda = v_\lambda^\mathrm{max}/ v_\lambda^\infty$ with $\gamma_\lambda$ (for those sectors contributing most), signalling 
a faster-than-linear scaling of the peak height $v_\lambda^\mathrm{max}$ with $\gamma_\lambda$.
Accordingly, the relative peak height $q_\lambda$ is largest for the bosonic sector $\lambda=(N)$,  which always has maximal purity $\gamma_{(N)}=1$  \cite{SM}.
This explains the sharp growth of $q$ observed in the inset 
of Fig.~\ref{fig:vk_vs_nu_pd} for $0 \lesssim \gamma \lesssim 0.2$, 
as the bosonic contribution to $v$ in Eq.~\eqref{eq:vk} surpasses contributions from non-bosonic sectors [Fig.~\ref{fig:vard1_vs_sectors}(a),(b)].
For $\gamma \gtrsim 0.4$, the bosonic contribution is dominant, as reflected by the convergence of $q$ towards $q_{(N)}$ [cf. inset Fig.~\ref{fig:vk_vs_nu_pd}].

We have thus shown that many-body coherences populated by the initial state leave a statistically robust imprint in the long-time fluctuations of few-particle observables.
The emergence of the chaotic phase induces the delocalization of the initial state in the eigenbasis, translating into an augmented contribution of coherences within the observable's energy bandwidth, and hence leading to the maximization of fluctuations.
This reflects the enhanced sensitivity of a quantum system's eigenstate structure (anchored in the underlying phase space's topological metamorphosis \cite{LesHouches})
at the chaos transition, which is inherited by single- as well as by many-particle transition amplitudes \cite{robbins_discrete_1989,seligman1994,schlagheck2019}.
While this fluctuation maximum is observed for any degree of particle distinguishability, it is significantly amplified as the particles become more indistinguishable, because of many-body interference (MBI) contributions stemming from the bosonic symmetry sector.
Therefore, full resolution of the particle-exchange symmetry sectors is indispensable to understand how  MBI is seeded by the spectral and eigenstate structure of a many-body quantum system, and to discern MBI's impact on the dynamics. Ultimately, this approach allows the discrimination of interaction-induced from entirely quantum (due to many-particle interferences) causes of dynamical complexity.

\begin{acknowledgments}
The authors thank Dominik Lentrodt for fruitful discussions.
The authors acknowledge support by the state of Baden-Württemberg through bwHPC (High Performance Computing, Baden-Württemberg), and funding by the Deutsche Forschungsgemeinschaft (DFG, German Research Foundation)---Grants No. INST 40/575-1 FUGG (JUSTUS 2 cluster) and No. 402552777.
E. G. C. acknowledges support from the Georg H. Endress Foundation.
E. B., L. P., and A. R. acknowledge support by Spanish MCIN/AEI/10.13039/501100011033 (Ministerio de Ciencia e Innovación/Agencia Estatal de Investigación) through Grant No. PID2020–114830GB-I00.
\end{acknowledgments}


\begin{thebibliography}{78}
\bibitem [{N1()}]{N1}%
  \BibitemOpen
  \href@noop {} {}\bibinfo {note} {Note that $v$ is independent of the chosen
  operator basis.}\BibitemShut {Stop}%
\end{thebibliography}

%

%
%


\newcommand{\TITLE}{Many-body interference at the onset of chaos}
\title{Supplemental Material \\[2mm] \TITLE}


\onecolumngrid
\begin{center}\large\bfseries Supplemental Material \\[2mm] \TITLE \end{center}
\vskip 4mm
\twocolumngrid
\renewcommand\theequation{S\arabic{equation}}
\renewcommand\thefigure{S\arabic{figure}}
\setcounter{figure}{0}
\setcounter{equation}{0}

\subsection{Distribution of internal states}

For the numerical simulations shown in Figs.~1 and 2 of the main text, we need to generate random internal states for each particle, so as to cover the transition from distinguishable to indistinguishable particles, quantified in terms of the purity $\gamma = \mathrm{tr} \rho^2$ of the reduced external state $\rho$, as uniformly as possible. The straightforward ansatz to generate random initial states, e.g., distributed according to the Haar measure on $\mathcal{H}\inint$, likely generates rather orthogonal states $\ket{\phi_i}$ for the particles and, thus, only samples the region of small $\gamma$.
To circumvent this problem, we employ the same two-step sampling procedure as used in Ref.~\cite{brunner_many-body_2022}:
We generate random pure internal states $\ket{\phi_i} = \sum_\sigma \phi_{i\sigma} \ket{\sigma}$ for the particle in mode $i$, where the states $\ket{\sigma}$ form a basis of $\mathcal{H}\inint$
(the dimension of the internal Hilbert space has to be larger than or equal to the number of particles).
To cover the vicinity of indistinguishable particles, we initialize a unit vector $\ket{e}\in\mathcal{H}\inint$ and perturb it by a random vector $\ket{g_i}$, with normally distributed real and imaginary parts of the components of $\ket{g_i}$, with zero mean and variance $\varepsilon$.
For sufficiently small $\varepsilon$, the resulting states $\ket{\phi_i}= \ket{e} + \ket{g_i}$, $i =1,\dots, L$, after renormalization, are almost parallel and the particles thus remain near-indistinguishable.
As $\varepsilon$ increases, the relative contribution of the constant vector $\ket{e}$ becomes negligible and we sample the unit sphere in $\mathcal{H}\inint$ uniformly in the limit $\varepsilon\gg1$.
In the second step, we employ a similar procedure in the neighborhood of distinguishable particles. For this we choose for each particle $i$ an orthogonal unit vector $\ket{e_i}\in\mathcal{H}\inint$ and, again, add a perturbation $\ket{g_i}$ with normally distributed components in $\mathbb{C}$, with zero mean and variance $\varepsilon$, followed by renormalization. For large $\varepsilon$, the contribution from $\ket{e_i}$ to $\ket{\phi_i}= \ket{e_i} + \ket{g_i}$ can be neglected, leading to uniform sampling of the unit sphere in $\mathcal{H}\inint$. As $\varepsilon$ becomes small, $\varepsilon \ll 1$, we sample states $\ket{\phi_i}$ close to perfectly distinguishable particles.

\begin{figure}
\centering
\includegraphics[width=0.48\textwidth]{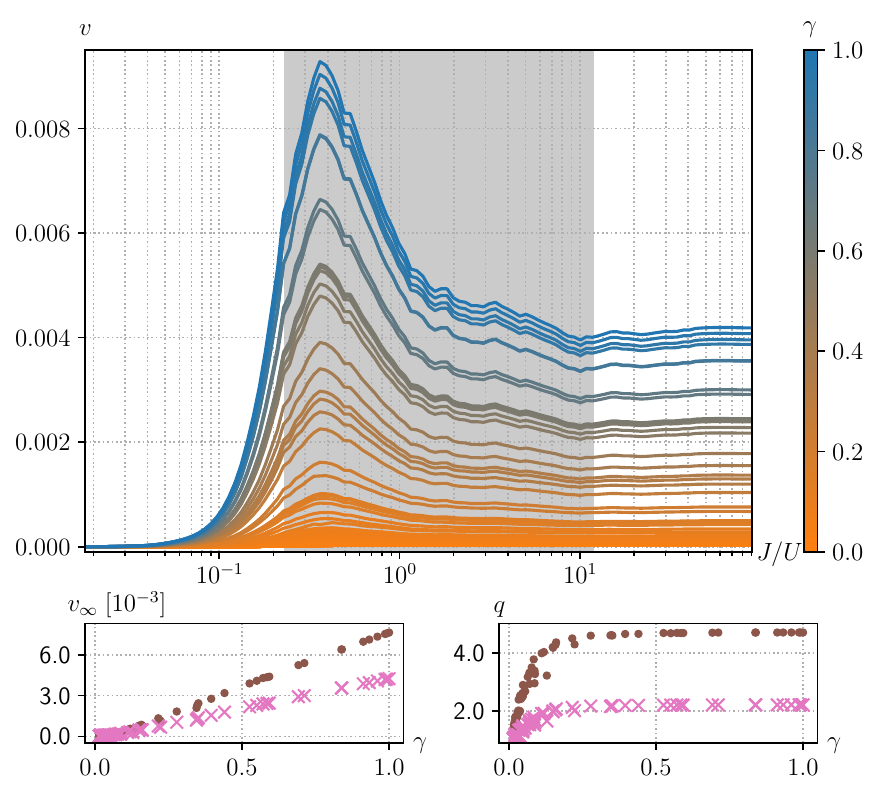}
\caption{Asymptotic temporal fluctuations $v$ of equally weighted two-point correlators according to Eq.~\eqref{eq:sm:cor_average}, as a function of $J/U$ for 100 random initial states of variable partial distinguishability, as quantified by the external state's purity $\gamma$ (color bar). The bottom panels compare the plateau value $v_{\infty}$ and the enhancement $q$ for $v$ as defined in Eq.~\eqref{eq:sm:cor_average} (crosses) to the results obtained for $v$ as defined in Eq.~(3) and already shown in the inset of Fig.~1 (dots), as functions of $\gamma$.}
\label{fig:vk_vs_nu_pd_kPcor}
\end{figure}

\subsection{Density correlation mean}

The operator average over a basis of $k$P observables [cf. {Eq.~(3)} of the main text] yields a statistically robust estimate of experimentally accessible $k$-point correlation measurements of the external modes.
To show this for the case $k=2$, we replace the average in Eq.~(3) by an average over all two-point density correlations
\begin{equation}\label{eq:sm:cor_average}
v = \frac{1}{L(L-1)} \sum_{i\neq j } \mathrm{Var}_t [N_i N_j] \,,
\end{equation}
where the operator $N_{i}=\sum_{\sigma=1}^s a_{i\sigma}^\dagger a_{i\sigma}$ counts the number of particles on site $i$, irrespective of their internal states.
Figure~\ref{fig:vk_vs_nu_pd_kPcor} shows $v$ for this average as a function of $J/U$ and of the 
particles' indistinguishability, quantified by $\gamma = \mathrm{tr} \rho^2$ (as in Fig.~1 of the main text).
We observe the same behavior as for the operator average, Eq.~(3). The average $v$ \eqref{eq:sm:cor_average} shows a peak for $J/U \approx 0.35$ (note a small shift to larger $J/U$ values in comparison to the operator average) followed by a decline to a plateau value $v_\infty$.
The dependence of $v_\infty$ and of the enhancement $q = v_\mathrm{max}/v_\infty$ on $\gamma$ is shown in the lower panels of Fig.~\ref{fig:vk_vs_nu_pd_kPcor} and compared to the results shown in the insets of Fig.~1.
We observe, up to a scaling factor, exactly the same, strictly monotonic increase of both quantities with $\gamma$ as in Fig.~1.
To resolve this constant scaling factor between both averages, one needs to divide Eq.~(3) by the dimensionality of the space of external two-particle observables. We omit this rather technical procedure here, since it is of no importance for our subsequent discussion.

\subsection{Schur-Weyl duality}
Since dynamics and measurements are restricted to the external dof only, we trace out the internal dof. The reduced external system is conveniently described by the $N$th tensor power of the external single-particle Hilbert space, $\mathcal{H}\inext^{\otimes N}$. On this space, the symmetric and the unitary groups, $\SN$ and $U(L)$, act according to $\pi \ket{m_1,\dots, m_N} \coloneqq \ket{m_{\pi^{-1}(1)}, \dots, m_{\pi^{-1}(N)}}$, $\pi\in\SN$, and $U \ket{m_1,\dots, m_N} \coloneqq U\ket{m_1} \otimes \dots \otimes U\ket{m_N}$, $U \in U(L)$, respectively. Schur-Weyl duality states that these two group actions are dual to each other \cite{Fulton2004}.
This implies that the external $N$-particle space decomposes into a direct sum of irreducible representations of $\SN$ and $U(L)$, $\mathcal{H}^\lambda_{\SN}$ and $\mathcal{H}^\lambda_{U(L)}$, i.e.,
\begin{equation}\label{eq:youngdecomp}
\mathcal{H}\inext^{\otimes N} = \bigoplus_{\lambda} \left(\mathcal{H}^\lambda_{\SN} \otimes \mathcal{H}^\lambda_{U(L)}\right)\,.
\end{equation}
The direct sum runs over Young diagrams (i.e., integer partitions of $N$), such as $\lambda = (N), (N-1,1), (N-2, 1,1),(N-2,2), \dots, (1,\dots, 1)$. While the reduced states of perfectly indistinguishable bosons only occupy the symmetric sector (which we therefore call the \textit{bosonic} sector), $\lambda = (N)$, the reduced states of partially distinguishable particles typically occupy all sectors, as {described} 
in the main text.

Equation \eqref{eq:youngdecomp} provides a convenient basis $\ket{\lambda, i, m} = \ket{\lambda, i} \otimes \ket{\lambda, m}$ of $\mathcal{H}\inext^{\otimes N}$, where $i$ and $m$ index basis states of the irreducible representations $\mathcal{H}^\lambda_{\SN}$ and $\mathcal{H}^\lambda_{U(L)}$, respectively. Each such basis state of $\mathcal{H}^\lambda_{\SN}$ and $\mathcal{H}^\lambda_{U(L)}$ corresponds to a standard, respectively semi-standard Young tableau of shape $\lambda$ \cite{Fulton2004,dufour_many-body_2020}. The number of standard and semi-standard Young tableaux, $\nu_\lambda$ and $d_\lambda$, are combinatorial in nature and can be calculated via hook length formulas \cite{Fulton2004}. 
The table below lists them for the considered case of $N = L = 6$.
\begin{center}
\begin{tabular}{ c | c c}
$\lambda$ & $\nu_\lambda$ & $d_\lambda$ \\ \hline
$(6)$ & 1 & 462 \\ 
$(5,1)$ & 5 & 1050 \\ 
$(4,2)$ & 9 & 1134 \\ 
$(4,1,1)$ & 10 & 840 \\ 
$(3,3)$ & 5 & 490 \\ 
$(3,2,1)$ & 16 & 896 \\ 
$(3,1,1,1)$ & 10 & 280 \\ 
$(2,2,2)$ & 5 & 175 \\ 
$(2,2,1,1)$ & 9 & 189 \\ 
$(2,1,1,1,1)$ & 5 & 35 \\ 
$(1,1,1,1,1,1)$ & 1 & 1 \\ 
\end{tabular}
\end{center}
Adding up the dimensions with their corresponding multiplicities leads to the total dimension $\sum_\lambda \nu_\lambda d_\lambda = L^N = 46656$ for the considered system.
Note that the dynamics of the system is exclusively described by the irreducible representations of the unitary group, $\mathcal{H}^\lambda_{U(L)}$. Hence, the numbers $\nu_\lambda$ of basis states $\ket{\lambda,i}$ of the irreducible representations of the symmetric group constitutes merely a multiplicity factor, and each symmetry sector $\lambda$ decomposes into $\nu_\lambda$ \textit{identical} blocks $\mathcal{H}^\lambda_{U(L)}$ of dimension $d_\lambda$. Since all blocks of one sector yield identical contributions to the dynamics, we can restrict the discussion to one of them for each sector, as done in Eq.~(4).

An initial state with well-defined external occupation numbers $N_i$ occupies all Young basis states $\ket{\lambda, m}$ corresponding to semi-standard Young tableaux of shape $\lambda$ compatible with the given occupation numbers. More precisely, these are exactly those Young tableaux that can be obtained by filling $N_i$ symbols `$i$' (for $i=1,\dots, L$) into the diagram $\lambda$ by following the rules that each column must be strictly increasing and each row must be non-decreasing. The number of such tableaux for given occupation numbers $\lbrace N_i \rbrace$ is given by the so called Kostka-number $\kappa_\lambda(\lbrace N_i \rbrace )$ \cite{Fulton2004}.
In case of the homogeneous initial state ($N_i = 1, i=1,\dots,L$) considered in Figs.~1 and 2 of the main text, semi-standard and standard Young diagrams are actually identical, leading to $\kappa_\lambda = \nu_\lambda$.
In the limit $J/U \rightarrow 0$, where the energy eigenstates approach the Young basis states, maximal localization is achieved with maximal inverse participation ratio $I_\lambda = \sum_\alpha | \rho^\lambda_{\alpha\alpha} |^2$, taking on a value larger than the inverse number of occupied Young basis states, i.e., $I_\lambda  \geq 1/\kappa_\lambda$ (the populations on the occupied Young basis states are typically not uniform).
This localization on a small number $\kappa_\lambda \sim 1$ of energy states for $J/U \rightarrow 0$, leads to a significant suppression of the numerator $\gamma_\lambda - I_\lambda$ of $R_\lambda$ [cf. Eq.~(5)] in this limit.
Note, moreover, that a lower tight bound of the purity $\gamma_\lambda$ for each sector is given by $1/\kappa_\lambda$, which is achieved for perfectly distinguishable particles. Since $\kappa_{(N)} = 1$, this implies maximal purity $\gamma_{(N)} = 1$ in the bosonic sector, independent of the particles' distinguishability.

\subsection{Derivation of Eq. (4) of the main text}

Under the assumption of a non-degenerate spectrum, the infinite time average of the time dependent expectation is given by
\begin{equation}\label{eq:sm:helper}
	\overline{\expv{O(t)}} = \sum_{\lambda,\alpha,\beta} p_\lambda \rho^\lambda_{\alpha\beta} O^\lambda_{\beta\alpha} \overline{ e^{it(E^\lambda_\alpha - E^\lambda_\beta)} } = \sum_{\lambda,\alpha} p_\lambda \rho^\lambda_{\alpha\alpha} O^\lambda_{\alpha\alpha} \,,
\end{equation}
with $\overline{\phantom{i}\ldots \phantom{i}}$ indicating the integration $1/T \int_0^T dt \dots$ for $T\rightarrow \infty$. Definitions of $\rho^\lambda_{\alpha\beta}, O^\lambda_{\alpha\beta}$ are given in the main text.
Integration over the exponential above yields $\delta_{\alpha\beta}$.
The second moment of the time signal $\expv{O(t)}$ is given by
\begin{equation}
	\overline{\expv{O(t)}^2} = \sum_{\substack{\lambda,\alpha,\beta \\ \tau,\gamma,\delta}} p_\lambda p_\tau \rho^\lambda_{\alpha\beta} O^\lambda_{\beta\alpha} \rho^\tau_{\gamma\delta}O^\tau_{\delta\gamma} \; \overline{e^{it(E^\lambda_\alpha - E^\lambda_\beta) + it(E^\tau_\gamma - E^\tau_\delta)}} \,.
\end{equation}
Assuming no degeneracies of levels and no gap-degeneracies between and within the $\lambda$-sectors, the time integration either decouples the sums over $\lambda$ and $\tau$ or enforces a $\delta_{\lambda\tau}$, leading to
\begin{equation}
\begin{split}
	\overline{\expv{O(t)}^2} &= \sum_{\lambda,\alpha, \tau,\gamma} p_\lambda p_\tau \rho^\lambda_{\alpha\alpha} O^\lambda_{\alpha\alpha} \rho^\tau_{\gamma\gamma} O^\tau_{\gamma\gamma} \\
	&\qquad + \sum_{\lambda,\alpha\neq\beta} p_\lambda^2 | \rho^\lambda_{\alpha\beta}|^2 | O^\lambda_{\alpha\beta}|^2 \,. 
\end{split}
\end{equation}
The first term is equal to the square of the time-averaged expectation $\overline{\expv{O(t)}}^2$, Eq.~\eqref{eq:sm:helper}. Subtracting this yields the variance
\begin{equation}
	\text{Var}_t [ O] = \sum_{\lambda,\alpha\neq\beta} p_\lambda^2  | \rho^\lambda_{\alpha\beta}|^2 | O^\lambda_{\alpha\beta}|^2 \,.
\end{equation}
Note that the variance is linear in $| O^\lambda_{\alpha\beta}|^2$, such that we can take the operator average over observables into the sum
\begin{equation}
	\widehat{\text{Var}_t [O]} = \sum_{\lambda,\alpha\neq\beta} p_\lambda^2  | \rho^\lambda_{\alpha\beta}|^2 \widehat{ | O^\lambda_{\alpha\beta}|^2} \,,
\end{equation}
which is Eq.~(4).

\subsection{Derivation of Eq. (5) of the main text}
We show the factorization of $v_\lambda$ into $R_\lambda A_\lambda$ according to Eq.~(4), under the considered statistical assumptions on the matrix elements $|\rho^\lambda_{\alpha\beta}|^2$ and $|O^\lambda_{\alpha\beta}|^2$, namely that $\rho^\lambda$ effectively occupies $d_\lambda^\mathrm{eff}$ energy levels, that the observable is banded with bandwidth $W_\lambda$, and that the matrix elements of state and observable are statistically independent. For convenience, we define $\delta^\rho_{\alpha}$ to be 1 on the $d_\lambda^\mathrm{eff}$ eigenstates (indexed by $\alpha$) closest in energy to the energy expectation $E = \text{tr}[\rho H] = 0$ of the considered homogeneous initial (Mott) state, and 0 elsewhere. Moreover let $\delta^O_{\alpha\beta}$ be 1 for $|\alpha - \beta| \leq W_\lambda$ and 0 for $|\alpha - \beta| > W_\lambda$. With this we can calculate
\begin{align*}
&\sum_{\alpha\neq \beta} |\rho^\lambda_{\alpha\beta}|^2 |O^\lambda_{\alpha\beta}|^2 = \sum_{\alpha\neq \beta} |\rho^\lambda_{\alpha\beta}|^2 |O^\lambda_{\alpha\beta}|^2  \delta^\rho_\alpha \delta^\rho_\beta \delta^O_{\alpha\beta} \\
&\quad = \frac{1}{\mathcal{N}} \sum_{\alpha\neq \beta} |\rho^\lambda_{\alpha\beta}|^2  \delta^\rho_\alpha \delta^\rho_\beta \delta^O_{\alpha\beta}  \sum_{\alpha'\neq \beta'}  |O^\lambda_{\alpha'\beta'}|^2  \delta^\rho_{\alpha'} \delta^\rho_{\beta'} \delta^O_{\alpha'\beta' } \\
&\quad = \frac{1}{\mathcal{N}} \frac{\mathcal{N} }{ (d_\lambda^\mathrm{eff})^2}\sum_{\alpha\neq \beta} |\rho^\lambda_{\alpha\beta}|^2  \delta^\rho_\alpha  \delta^\rho_\beta \frac{\mathcal{N} }{d_\lambda W_\lambda} \sum_{\alpha'\neq \beta'}  |O^\lambda_{\alpha'\beta'}|^2   \delta^O_{\alpha'\beta'} \\
&\quad = \frac{\min( d_\lambda^\mathrm{eff}, W_\lambda )}{d_\lambda d_\lambda^\mathrm{eff} W_\lambda}  \sum_{\alpha\neq \beta} |\rho^\lambda_{\alpha\beta}|^2   \sum_{\alpha'\neq \beta'}  |O^\lambda_{\alpha'\beta'}|^2 \\
&\quad = \frac{1}{d_\lambda \max( d_\lambda^\mathrm{eff}, W_\lambda ) }  \sum_{\alpha\neq \beta} |\rho^\lambda_{\alpha\beta}|^2   \sum_{\alpha'\neq \beta'}  |O^\lambda_{\alpha'\beta'}|^2  \,,
\end{align*}
with $\mathcal{N} = d_\lambda^\mathrm{eff} \min(d_\lambda^\mathrm{eff},W_\lambda)$. The second equality exploits the statistical independence of the matrix elements of state and observable. The sums in line two run over a subset of indices $\alpha\neq \beta$ of size $\mathcal{N}$. The sums in line three run over $(d_\lambda^\mathrm{eff})^2$ and $d_\lambda W_\lambda$ indices, respectively. For this we correct by the introduced fractions in line three.

\begin{figure}
\centering
\includegraphics[width=0.48\textwidth]{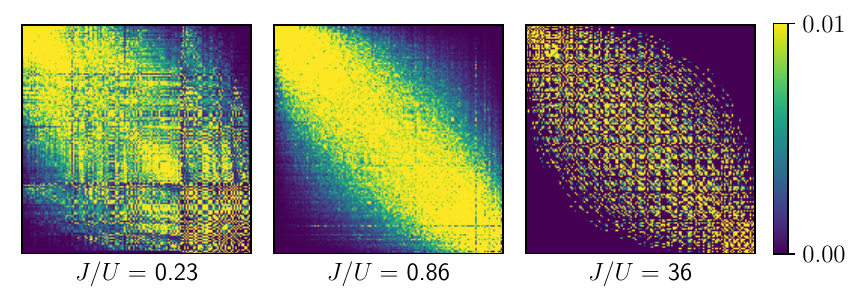}
\caption{Matrix plots of the operator averaged two-particle observable $\widehat{| O^\lambda_{\alpha\beta} |^2}$ [cf. {Eq.~(3)} of the main text], for three exemplary values of $J/U$, in the totally symmetric (bosonic) sector, and for $N=L = 6$.}
\label{fig:matrix_plots}
\end{figure}

\subsection{Banded structure of the observable in the eigenstates}

Figure~\ref{fig:matrix_plots} shows matrix plots of the operator averaged two-particle observable $\widehat{| O^\lambda_{\alpha\beta} |^2}$ [cf. {Eq.~(3)} of the main text] for three exemplary values of $J/U$ in the totally symmetric (bosonic) sector, for a system with 
$N = L = 6$.
For small $J/U = 0.23$ the observable does not yet develop a prominent band structure. However, in this parameter regime, the observable does not play a major role and the fluctuations $v$ [cf. Eq.~(4) of the main text] are dominated by the strong localization of the initial state on only a small energy window. The clearest uniform band structure is observed for intermediate $J/U = 0.86$. For large $J/U$, beyond the chaotic domain, a banded structure persists, however, the band is not uniformly occupied anymore.

\subsection{Contribution $A_\lambda$ of the observable {in Eq.~(5)} of the main text}

\begin{figure}
\centering
\includegraphics[width=0.48\textwidth]{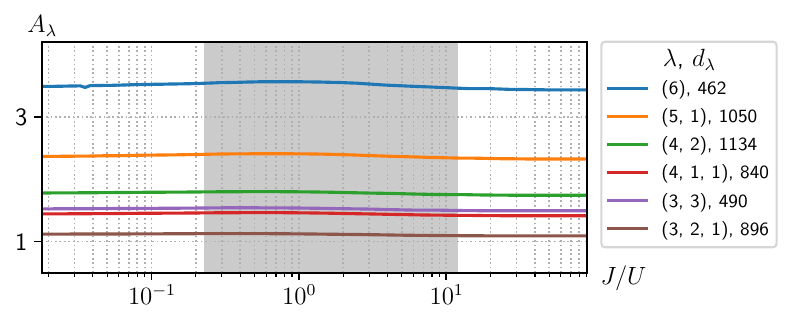}
\caption{The contribution $A_\lambda$ of the observable to {Eq.~(5)} in the main text for the largest six symmetry sectors as a function of $J/U$. $A_\lambda$ is almost constant and of order of one.}
\label{fig:A_lambda_plot}
\end{figure}

Figure~\ref{fig:A_lambda_plot} shows $A_\lambda$ [cf. {Eq.~(5)} in the main text] in the largest six symmetry sectors as a function of the control parameter $J/U$. A system with $N = L = 6$ is considered. As described in the main text, $A_\lambda$ is almost constant as a function of $J/U$, and of order of one in all shown sectors.

\subsection{Calculation of the averaged matrix elements $\widehat{|O^\lambda_{\alpha\beta} |^2}$ of the observable}

Here we derive the expression of $\widehat{|O^\lambda_{\alpha\beta} |^2}$ in terms of the Hilbert-Schmidt inner products of particle-reduced energy eigenstates, as given in the footnote [68] of the main text (and implicitly also appearing in Eqs.~(4) and (5)).
The $N$th tensor power of $\mathcal{H}\inext^{\otimes N}$ allows for a bipartition $\mathcal{H}\inext^{\otimes N} = \mathcal{H}\inext^{\otimes k} \otimes \mathcal{H}\inext^{\otimes (N-k)}$ into the reduced $k$P space and the remainder of the system.
Let $\lbrace O^{(k)}_i \rbrace_{i=1}^{d_\mathrm{OP}}$ be an orthonormal basis of the operator space of Hermitian $k$P operators which, together with the Hilbert-Schmidt inner product, forms a finite-dimensional Hilbert space of dimension $d_\mathrm{OP}$.
For pairs of eigenstates $\ket{E_\alpha}$ we define reduced operators $\varepsilon^{(k)}_{\alpha\beta} = \text{tr}_{(N-k)} [\op{E_\alpha} {E_\beta } ]$, which
for $\alpha = \beta$ are just the $k$P reduced density operators of the eigenstate \cite{brunner_many-body_2019}.
For brevity we set $A=(k), B=(N-k)$ and calculate
\begin{align*}
\widehat{|O_{\alpha\beta} |^2} &= \sum_{i=1}^{d_\mathrm{OP}} \text{tr} \big[ O^A_i \op{E_\beta}{E_\alpha} \big] \text{tr} \big[ O^A_i \op{E_\alpha}{E_\beta} \big] \\
&= \binom{N}{k}^2 \;\, \sum_{i=1}^{d_\mathrm{OP}} \text{tr}_A \big[ O^A_i \varepsilon^A_{\beta\alpha} \big] \text{tr}_A \big[ O^A_i \varepsilon^A_{\alpha\beta} \big] \\
&= \binom{N}{k}^2 \; \text{tr}_A \big[ \varepsilon^A_{\beta\alpha} \varepsilon^A_{\alpha\beta} \big] = \binom{N}{k}^2 \; \text{tr}_B \big[ \varepsilon^B_{\alpha\alpha} \varepsilon^B_{\beta\beta} \big] \,,
\end{align*}
where the third equality uses the fact that the $O^{(k)}_i$ are an orthonormal basis. The proportionality factor $\binom{N}{k}$ accounts for the fact that the (implicit) extension of $O_i^{(k)}$ to the full $N$P space $\mathcal{H}\inext^{\otimes N}$ is not normalized, more precisely $\text{tr} \big[ O^A_i \op{E_\beta}{E_\alpha} \big] = \binom{N}{k} \text{tr}_A \big[ O^A_i \varepsilon^A_{\beta\alpha} \big]$ \cite{brunner_many-body_2019}.
The last equality can be shown using the Schmidt decomposition $\ket{E_\alpha} = \sum_i c^\alpha_i \ket{\phi^A_i, \phi^B_i}$, which allows us to calculate
\begin{align*}
&\text{tr}_A \big[ \varepsilon^A_{\beta\alpha} \varepsilon^A_{\alpha\beta} \big] \\
&= \text{tr}_A \Big[ \text{tr}_B\Big[ \sum_{ij} c^\beta_i c^\alpha_j \op{\phi^A_i \phi^B_i}{\phi^A_j, \phi^B_j} \Big] \\
&\qquad\qquad \times \text{tr}_B\Big[ \sum_{k\ell} c^\alpha_k c^\beta_\ell \op{\phi^A_k \phi^B_k}{\phi^A_\ell, \phi^B_\ell} \Big] \Big] \\
&= \sum_{ijk\ell} c^\beta_i c^\alpha_j c^\alpha_k c^\beta_\ell \text{tr}_A \Big[ \text{tr}_B\Big[ \op{\phi^A_i \phi^B_i}{\phi^A_j, \phi^B_j} \Big] \\
&\qquad\qquad \times \text{tr}_B\Big[ \op{\phi^A_k \phi^B_k}{\phi^A_\ell, \phi^B_\ell} \Big] \Big] \\
&= \sum_{ijk\ell} c^\beta_i c^\alpha_j c^\alpha_k c^\beta_\ell \expv{\phi_j^B | \phi_i^B} \expv{\phi_\ell^B | \phi_k^B} \expv{\phi_j^A | \phi_k^A} \expv{\phi_\ell^A | \phi_i^A} \\
&\qquad\qquad = \text{tr}_B \big[ \varepsilon^B_{\alpha\alpha} \varepsilon^B_{\beta\beta} \big] \,.
\end{align*}

\end{document}